\documentclass{article}
\usepackage{spconfa4,amsmath,graphicx}
\usepackage{amsfonts}
\usepackage[nolist]{acronym}
\usepackage{comment}
\usepackage{xcolor} % Necessario per definire i colori
\usepackage{hyperref}
\usepackage{cleveref}
\usepackage{sidecap}
\usepackage{subcaption}

\usepackage{bm,cite}

\title{Learning-based Physics-Constrained Neural Kernel for Sound Field Estimation With Source-Position-Dependent Directional Weighting}
\name{Mattia Marella$^{1,2}$~\sthanks{This work was performed while Mattia Marella was an intern at the National Institute of Informatics. \\This work was supported by JST FOREST Program under Grant Number JPMJFR216M and ”Strategic Programs” grant from ROIS (Research Organization of Information and Systems).} \quad and \quad Shoichi Koyama$^1$}
\address{$^1$National Institute of Informatics, Tokyo, Japan, \quad $^2$ University of Ferrara, Ferrara, Italy}
\begin{document}
  \begin{acronym}
\acro{PIML}{Physics-Informed Machine Learning}
\acro{PDE}{Partial Differential Equation}
\acro{PINN}{Physics-Informed Neural Network}
\acro{RKHS}{Reproducing Kernel Hilbert Space}
\acro{BEM}{Boundary Element Method}
\acro{ESM}{Equivalent Source Method}
\acro{CS}{Compressed Sensing}
\acro{ATF}{Acoustic Transfer Function}
\acro{LSD}{Log-Spectral Distortion}
\acro{NMSE}{Normalized Mean Squared Error}
\acro{RIR}{Room Impulse Response}
\acro{SNR}{Signal-to-Noise Ratio}
\acro{AWGN}{Additive White Gaussian Noise}
\acro{KRR}{Kernel Ridge Regression}
\acro{FFT}{Fast Fourier Transform}
\acro{HPC}{High-Performance Computing}
\acro{LOO}{Leave-One-Out}
\acro{RFF}{Random Fourier Features}
\end{acronym}%

\ninept
\maketitle
\begin{abstract}
A learning-based physics-constrained neural kernel for sound field estimation is proposed. Sound field estimation aims to estimate the spatial distribution of an acoustic field from a discrete set of microphone measurements, which have a wide range of applications. Among existing sound field estimation methods, kernel-regression-based methods offer a flexible and principled framework for incorporating physical constraints and allow inference through linear operation. It is also possible to adapt the kernel function to the target acoustic environment by representing the directional weighting function as an implicit neural representation (INR) and optimizing hyperparameters using measurements. However, the kernel function is generally optimized for single snapshot measurements of the microphones, which can lead to strong overfitting and poor generalization. We propose a source-position-dependent INR for the directional weighting function, enabling the kernel function to capture common directional patterns and to generalize to unseen source positions in the target acoustic environment. Experimental results indicate that our proposed method outperforms the snapshot-based method by estimating a directional weighting function that matches the directivity of the target sound field. 
\end{abstract}
%

% ========================================================

\section{Introduction}
\label{sec:intro}
%\vfill

Sound field estimation/reconstruction/interpolation is a fundamental task in audio signal processing and machine learning, aiming to estimate the spatial distribution of an acoustic field from a discrete set of sensor (microphone) observations. It can be applied to a wide variety of downstream tasks, for example, acoustic imaging~\cite{Maynard:JASA1985}, room acoustic analysis~\cite{Park:JASA_J_2005}, and spatial audio reproduction and control~\cite{Poletti:J_AES_2005,Koyama:IEEE_ACM_J_ASLP2021}. 

There have been many studies on sound field estimation~\cite{Ueno:FnT_SP2025}. One of the most widely used techniques is the basis-expansion-based method, which is based on the representation of the sound field as a linear combination of predefined basis functions, such as plane waves, spherical wave functions, and equivalent point sources~\cite{Williams:FourierAcoust,Poletti:J_AES_2005}. The expansion coefficients are then estimated from the microphone measurements using the least squares method. Sparse regularization techniques have also been applied to promote sparsity in the coefficient space to improve reconstruction accuracy~\cite{Antonello:IEEE_ACM_J_ASLP2017,Koyama:IEEE_J_JSTSP2015,Bertin:CSAbook2015}. Kernel-regression-based methods generalize the (finite-dimensional) basis-expansion-based method to an infinite-dimensional basis expansion, thereby enabling the estimates to be constrained to the solution of the governing equation (wave and Helmholtz equations)~\cite{Ueno:IEEE_J_SP2021,Brunnstroem:IEEE_J_ASLPRO2026}. 

In recent years, neural network (NN)-based methods have attracted attention because of their high representational power and interpolation capabilities~\cite{Lluis:JASA2020, Luo:NIPS2022, Pezzoli:Sensors2022, Shigemi:IWAENC2022, Miotello:ICASSP2024}. Physics-informed (penalized and constrained) approaches have also been investigated in the NN-based methods to avoid overfitting and increase the interpretability~\cite{Pezzoli:ForumAcusticum2023, Ribeiro:IEEE_ACM_J_ASLP2024, Karakonstantis:JASA2024, Olivieri:EURASIP2024, Koyama:IEEE_M_SP2025}.  

Among these current sound field estimation methods, the kernel-regression-based methods are particularly practical for applications requiring real-time processing, as they satisfy physical constraints while enabling inference through linear operation. To design physics-constrained kernel functions, the Herglotz wave function~\cite{Colton:InvAcoust_2013}, which is equivalent to a plane wave expansion, is used. To incorporate prior information regarding the directivity of the target sound field, a directional weighting function can be introduced into the Herglotz wave function~\cite{Ueno:IEEE_J_SP2021}. Although a fixed unimodal function or its linear combination has been used as the directional weighting function, they are unable to fully capture the characteristics of the target acoustic environment. Ribeiro~et~al.~\cite{Ribeiro:IEEE_ACM_J_ASLP2024} proposed a method for adapting physics-constrained kernel functions to the target acoustic environment by introducing implicit neural representations (INR)~\cite{Sitzmann:NeurIPS2020} and optimizing the hyperparameters based on the measurement data, which is referred to as \textit{physics-constrained neural kernel}. However, since the kernel function is adapted only to a single snapshot of measurement data, updating it requires significant computational effort for fine-tuning the hyperparameters, resulting in strong overfitting and poor generalization performance. 

We propose a learning-based physics-constrained neural kernel, which enables adapting the kernel function to the target acoustic environment based on a set of training data. The training data is assumed to be obtained in advance through numerical simulations and/or practical measurements of acoustic transfer functions (ATFs). The directional weighting function is designed as an INR depending on the source positions. Thus, the source-dependent characteristics of the target acoustic environment are expected to be captured by the training data, and the directional weighting function optimal to the given source positions can be obtained without finetuning. Experimental evaluations are performed to compare the proposed method with the snapshot-based physics-constrained neural kernel. 

% ========================================================
\section{Problem Formulation}
\label{ssec:problem}

Let $\Omega \subset \mathbb{R}^3$ be a simply-connected, source-free region of interest. The acoustic pressure field of angular frequency $\omega$ at position $\mathbf{x} \in \Omega$ is denoted by $u(\mathbf{x}, \omega)$, which satisfies the homogeneous Helmholtz equation:
\begin{equation}
    \nabla^2 u(\mathbf{x}, \omega) + k^2 u(\mathbf{x}, \omega) = 0, \quad \forall \mathbf{x} \in \Omega,
    \label{eq:helmholtz}
\end{equation}
where $k = \omega / c$ is the wavenumber and $c$ is the speed of sound. Sound field estimation aims to reconstruct $u(\mathbf{x},\omega)$ throughout $\Omega$ from a finite set of measurements $\{s_m(\omega)\}_{m=1}^M$ collected at known microphone positions $\{\mathbf{x}_m\}_{m=1}^M \subset \Omega$. Hereafter, we omit the frequency dependence for simplicity.

In the learning-based approach to sound field estimation, a set of ATFs from the sources at known positions $\{\mathbf{y}_j \}_{j=1}^J \subset \mathbb{R}^3 \backslash \Omega$ to the target positions $\{\mathbf{x}_i \}_{i=1}^I \subset \Omega$ in a static acoustic environment is assumed to be available, which is denoted as $\{h(\mathbf{x}_i, \mathbf{y}_j)\}_{i,j}$. They could be obtained by numerical simulations and/or practical measurements of the target environment. The goal is to estimate $u$ for unseen sources at given positions from the measurements $\bm{s}=[s_1, \ldots, s_M]^{\mathsf{T}}$, exploiting the features of the target acoustic environment extracted from the training ATFs. 

\section{Related Work}
\label{sec:prior}

We here briefly revisit the kernel-regression-based sound field estimation methods, which can be regarded as a generalization of finite-dimensional basis-expansion-based methods. They allow for interference through linear operation and require few parameters that have to be set empirically, such as the number of basis functions. 

\subsection{Kernel Regression for Sound Field Estimation}

The kernel-regression-based methods provide a flexible and principled framework for incorporating physical constraints. The problem to be solved is formulated as
\begin{equation}
    \hat{u} = \arg\min_{u \in \mathcal{H}}
    \sum_{m=1}^{M} |s_m - u(\mathbf{x}_m)|^2 + \lambda \|u\|_{\mathcal{H}}^2,
    \label{eq:KRR_functional}
\end{equation}
where $\mathcal{H}$ is a \ac{RKHS} of functions that satisfy the Helmholtz equation \eqref{eq:helmholtz}, and $\lambda > 0$ is a regularization parameter. The optimal solution can be expressed in closed form using the representer theorem~\cite{Scholkopf:COLT2001} as
\begin{equation}
    \hat{u}(\mathbf{x}) = \bm{\kappa}(\mathbf{x})^{\mathsf{T}} \left( \mathbf{K} + \lambda \mathbf{I} \right)^{-1} \mathbf{s},
    \label{eq:representer}
\end{equation}
where $\bm{\kappa}(\mathbf{x}) = [\kappa(\mathbf{x}, \mathbf{x}_1), \ldots, \kappa(\mathbf{x}, \mathbf{x}_M)]^{\mathsf{T}}$ is the vector of kernel evaluations and $\mathbf{K}$ is the Gram matrix with entries $\mathbf{K}_{ij} = \kappa(\mathbf{x}_i, \mathbf{x}_j)$.

The choice of \ac{RKHS} $\mathcal{H}$ as well as the kernel function $\kappa$ is crucial for the performance of kernel-regression-based methods. Ueno~et~al.~\cite{Ueno:IEEE_J_SP2021} proposed a kernel function based on Herglotz wave functions, which is a weighted integral of plane waves over the unit sphere, expressed as
\begin{equation}
    \kappa(\mathbf{x}, \mathbf{x}') = \frac{1}{4\pi} \int_{\mathbb{S}^2} \mathrm{e}^{\mathrm{j}k \bm{\eta} \cdot (\mathbf{x} - \mathbf{x}')} w(\bm{\eta}) \mathrm{d}\bm{\eta},
    \label{eq:kernel_continuous}
\end{equation}
where $w: \mathbb{S}^2 \to \mathbb{R}_{\ge 0}$ is a directional weighting function that enhances the directional sensitivity of the kernel. In \cite{Ueno:IEEE_J_SP2021}, the unimodal weighting function based on the von Mises--Fisher distribution~\cite{Mardia:DirStat} is proposed, which admits a closed-form expression for the kernel. 

\subsection{Neural Kernels for Sound Field Estimation}

The unimodal directional weighting function in \cite{Ueno:IEEE_J_SP2021} is limited in its ability to capture complex directional patterns of sound fields, especially in reverberant environments. To address this limitation, Ribeiro~et~al.~\cite{Ribeiro:IEEE_ACM_J_ASLP2024} proposed to model the directional weighting function $w$ as a sum of directed and residual components. The directed component is a superposition of von Mises--Fisher distributions for capturing direct sound and early reflections, and the residual component is an INR consisting of multilayer perceptrons (MLPs) for capturing late reverberation. The model parameters are jointly optimized by the gradient-descent-based method to minimize the reconstruction error. However, a single snapshot measurement is used to optimize the model parameters, which may lead to overfitting and poor generalization performance. 

\section{Proposed Method}
\label{sec:method}

\begin{figure}[t!]
    \centering
    \includegraphics[width=0.42\textwidth]{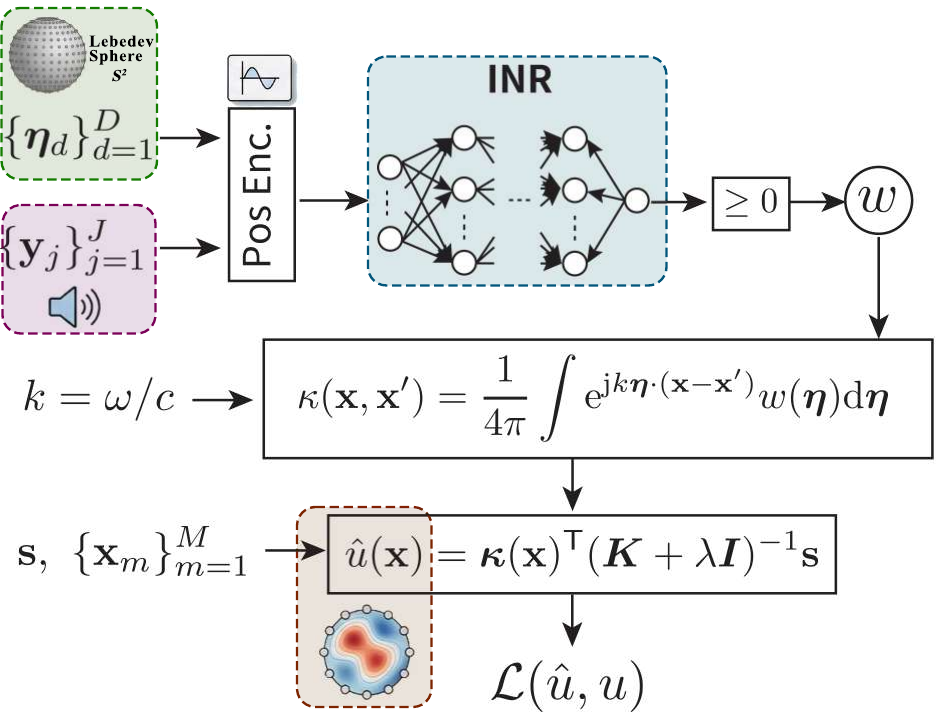}
        \caption{Conceptual diagram of the proposed physics-constrained neural kernel. The training pipeline combines the finely-discretized directions on $\mathbb{S}^2$ with a source-dependent neural modulation that produces positive directional weights $w$, which are used to construct the kernel function and reconstruct the sound field.}
        \label{fig:pipeline_a}
\end{figure}

\begin{figure}[t!]
    \centering
        \includegraphics[width=0.28\textwidth]{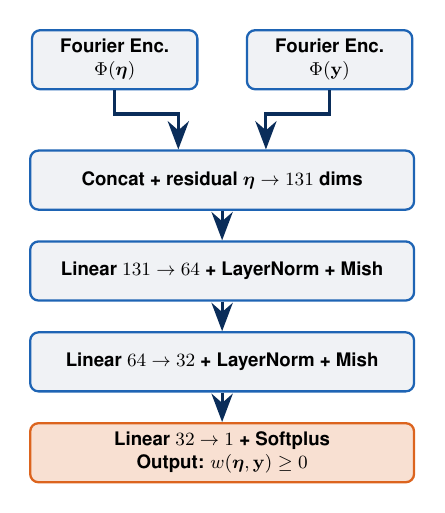}
        \caption{Proposed NN architecture. The INR takes RFM embeddings of the direction $\bm{\eta}$ and source position $\mathbf{y}$, concatenated with a residual connection, and maps them through an MLP to a non-negative weight via a Softplus output.}
        \label{fig:pipeline_b}
\end{figure}

We propose a learning-based neural kernel for sound field estimation, which extends the physics-constrained neural kernel framework of \cite{Ribeiro:IEEE_ACM_J_ASLP2024} to a multi-source training paradigm. Whereas the current method fits independent kernel instances for each source position using snapshot measurements, the proposed approach learns a directional weighting function shared across multiple sources, enabling the model to capture common directional patterns and to generalize to unseen source positions. 

\subsection{Learning-based Physics-Constrained Neural Kernel}
\label{ssec:kernel}

We represent the directional weighting function $w$ in \eqref{eq:kernel_continuous} as an INR with the input of $\bm{\eta}$ and source position $\mathbf{y}\in\mathbb{R}^3 \backslash \Omega$ as
\begin{equation}
w(\bm{\eta}) = \mathrm{INR}(\bm{\eta}, \mathbf{y}; \bm{\theta}),
\label{eq:inr_dir}
\end{equation}
where $\bm{\theta}$ denotes the NN weights. Thus, the source-position-dependent directional weighting function can be constructed. The kernel function is computed based on \eqref{eq:kernel_continuous}; therefore, the estimated field by using this kernel function, as in \eqref{eq:representer}, still satisfies the Helmholtz equation. (see Fig.~\ref{fig:pipeline_a})

Although the directional weighting function in \eqref{eq:inr_dir} is a representation of a continuous function with respect to $\bm{\eta}$, the kernel function \eqref{eq:kernel_continuous} has to be obtained by numerical integration, as obtaining its analytical expression is normally intractable. Owing to the INR, we can compute the integration by finely discretizing the unit sphere surface $\mathbb{S}^2$. The discrete directions are denoted by $\{\bm{\eta}_d\}_{d=1}^D$. We applied the Lebedev quadrature~\cite{Lebedev:DM1999} for sampling on $\mathbb{S}^2$ and the discrete directions $\{\bm{\eta}_d\}_{d=1}^D$ are fixed during the training; thus, the translational invariance of the kernel function $\kappa(\mathbf{x},\mathbf{x}^{\prime})=\kappa(\mathbf{x}-\mathbf{x}^{\prime})$ is preserved.

The kernel function based on the INR is also used in the previous method~\cite{Ribeiro:IEEE_ACM_J_ASLP2024} for the residual component. We additionally include the source-position dependence $\mathbf{y}$ in the directional weighting \eqref{eq:inr_dir}. Thus, it is expected to enable learning of optimal directional weighting functions based on the source positions $\mathbf{y}$ from the training data. During inference, it will be possible to obtain optimal directional weighting functions even for unseen sources at given positions only by the forward propagation of the INR. On the contrary, the previous method \cite{Ribeiro:IEEE_ACM_J_ASLP2024} uses only the direction $\bm{\eta}$ as input; therefore, the directional weighting function is optimized solely with the snapshot observations $\bm{s}$, and fine-tuning of hyperparameters is required for each measurement.

\subsection{Proposed Network Architecture}

The source-position-dependent INR, $\mathrm{INR}: (\bm{\eta},\mathbf{y}) \to \mathbb{R}_{\ge 0}$, is constructed as shown in Fig.~\ref{fig:pipeline_b}. Both the direction $\bm{\eta} \in \mathbb{S}^2$ and the source position $\mathbf{y}\in\mathbb{R}^3\backslash\Omega$ are first lifted into high-dimensional embeddings via Random Fourier Features (RFF)~\cite{Tancik:NeurIPS2020}:
\begin{equation}
    \Phi(\mathbf{z}) = \left[\sin(2\pi \mathbf{B}\mathbf{z}),\, \cos(2\pi \mathbf{B}\mathbf{z})\right],
    \label{eq:rff}
\end{equation}
where $\bm{z}$ is $\bm{\eta}$ or $\mathbf{y}$, and $\mathbf{B}\in\mathbb{R}^{32 \times 3}$ is a fixed Gaussian random matrix whose bandwidth is scaled proportionally to $k$, increasing representational capacity with frequency. These two embeddings are concatenated with a residual connection of the raw direction $\bm{\eta}$, followed by a three-layer MLP with Layer Normalization~\cite{Ba2016Layer} and Mish activations~\cite{Misra2019Mish}. The final output is enforced to be non-negative by using Softplus activation function~\cite{Dugas2000Incorporating}. Thus, the positive semi-definiteness of the kernel function $\kappa$ as well as the Gram matrix $\mathbf{K}$ is guaranteed. 

The model is trained by using a set of ATFs $\{h(\mathbf{x}_i,\mathbf{y}_j)\}_{i,j}$ from $J$ sources to $I$ evaluation points in a single fixed room. The loss function is the normalized mean square error (NMSE) of the reconstructed pressures at the evaluation points. 
\begin{equation}
\mathcal{L} = \frac{\sum_{i,j} |\hat{u}(\mathbf{x}_i; \mathbf{y}_j, \bm{\theta}) - u(\mathbf{x}_i; \mathbf{y}_j)|^2}{\sum_{i,j} |u(\mathbf{x}_i; \mathbf{y}_j)|^2},
\end{equation}
where $\hat{u}$ is the estimated pressures from the measurements $\mathbf{s}$ by \eqref{eq:representer} using the kernel function \eqref{eq:kernel_continuous} with the directional weighting \eqref{eq:inr_dir}. 

% =============================================================
%  EXPERIMENTS
% =============================================================
\section{Experiments}
\label{sec:experiments}

\subsection{Experimental Setup}

To construct a dataset, synthetic room impulse responses (RIRs) are generated by using the image source method~\cite{Allen:JASA1979}. The simulated room is $4.0~\mathrm{m} \times 6.0~\mathrm{m} \times 3.0~\mathrm{m}$ shoebox geometry, and the target reverberation time is set to $T_{60}=200~\mathrm{ms}$. The target region $\Omega$ is a spherical region of radius $0.5~\mathrm{m}$ centered at $(2.0~\mathrm{m}, 3.0~\mathrm{m}, 1.5~\mathrm{m})$ with the coordinate origin at the bottom corner of the room. The target region is discretized into $I=1331$ points every $0.1~\mathrm{m}$ to obtain the evaluation points. $100$ point sources are randomly placed inside the room, outside $\Omega$ and at a distance of $0.2~\mathrm{m}$ from the walls and $\Omega$. The sources are split into training, validation, and test sets at the ratios of 80\%, 10\%, and 10\%, respectively. The sampling frequency is $4800~\mathrm{Hz}$ and sound speed is $c=343~\mathrm{m/s}$. The complex ATFs are obtained by FFT of size $2048$. 

Omni-directional microphones are placed on two concentric spherical layers of radius $0.5~\text{m}$ and $0.49~\text{m}$. A total of $M$ microphones are distributed equally between these layers, with $M/2$ sensors per shell. The microphones on both layers share the same angular orientations $(\theta_m, \phi_m)$, effectively forming $M/2$ radially aligned pairs to avoid the forbidden frequency problem~\cite{Williams:FourierAcoust}. The points are distributed according to a spherical $t$-design~\cite{Chen:SIAM_NumerAnal2006}.

We compare the proposed learning-based neural kernel (LB-NK) with the snapshot-based neural kernel (SB-NK) and the fixed kernel using uniform weighting (Uniform). SB-NK is based on the same network architecture as LB-NK for constructing the directional weighting function $w$, but the source-dependent components are removed. Thus, the network parameters are optimized only for the snapshot observation $\mathbf{s}$. The optimization is performed with the Adam optimizer~\cite{Kingma2014Adam} with ReduceLROnPlateau scheduler at an initial learning rate of $10^{-3}$. The early-stopping criterion for LB-NK is based on the validation loss $\mathcal{L}$. Since SB-NK does not have the validation dataset, the number of epochs for optimizing the network parameters is fixed at $300$. Uniform does not require iterative optimization, and $\hat{u}$ is obtained by \eqref{eq:representer} with the fixed kernel function using $w=1.0$. The reconstructed pressure distribution by each method is evaluated by the test loss $\mathcal{L}$. 

\begin{figure}[t!]
    \centering
    \includegraphics[width=0.7\columnwidth]{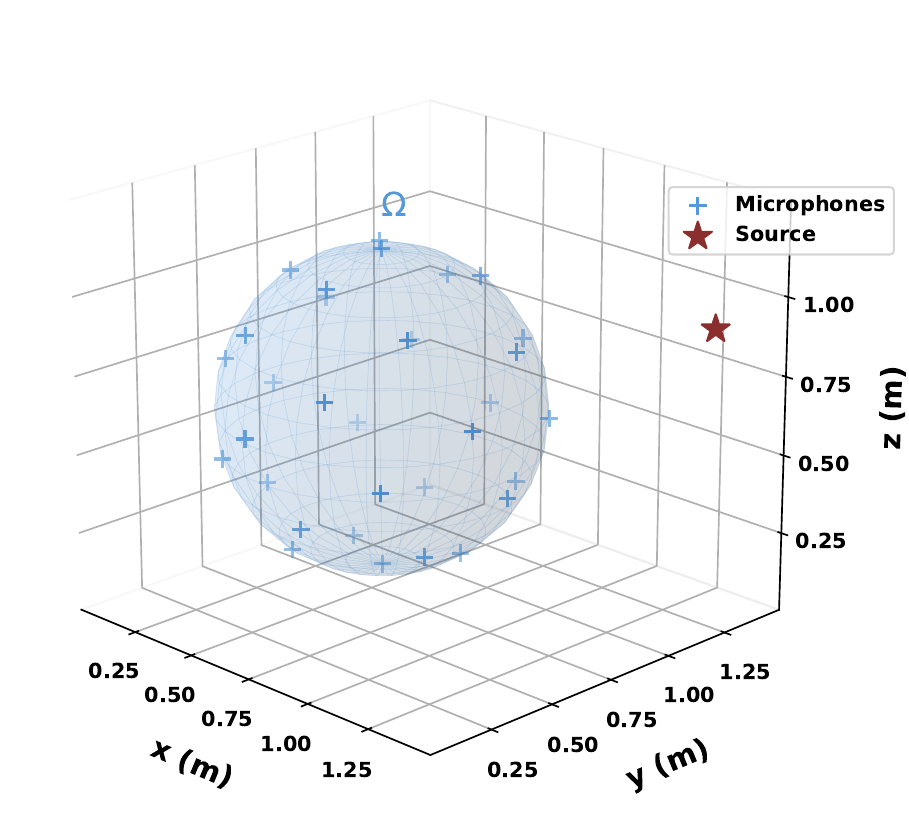}
    \caption{Experimental setup. Omnidirectional microphones are placed on two spherical layers of radii $0.50~\mathrm{m}$ and $0.49~\mathrm{m}$. $I=1331$ target positions are regularly arranged inside the inner spherical shell.}
    \label{fig:room}
\end{figure}

\subsection{Results}
\label{sec:results}

\begin{figure}[t!]
    \centering
    \includegraphics[width=0.95\columnwidth]{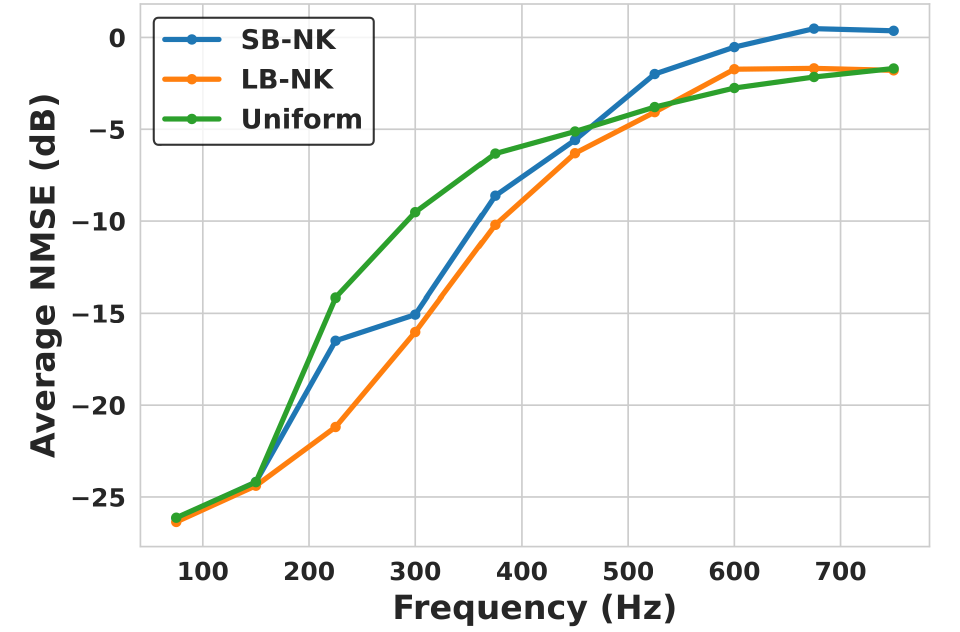}
    \caption{Average NMSE of SB-NK, LB-NK, and Uniform with respect to the frequency.}
    \label{fig:nmse_18_average}
\end{figure}

\begin{figure}[t!]
    \centering
    \includegraphics[width=1.0\columnwidth]{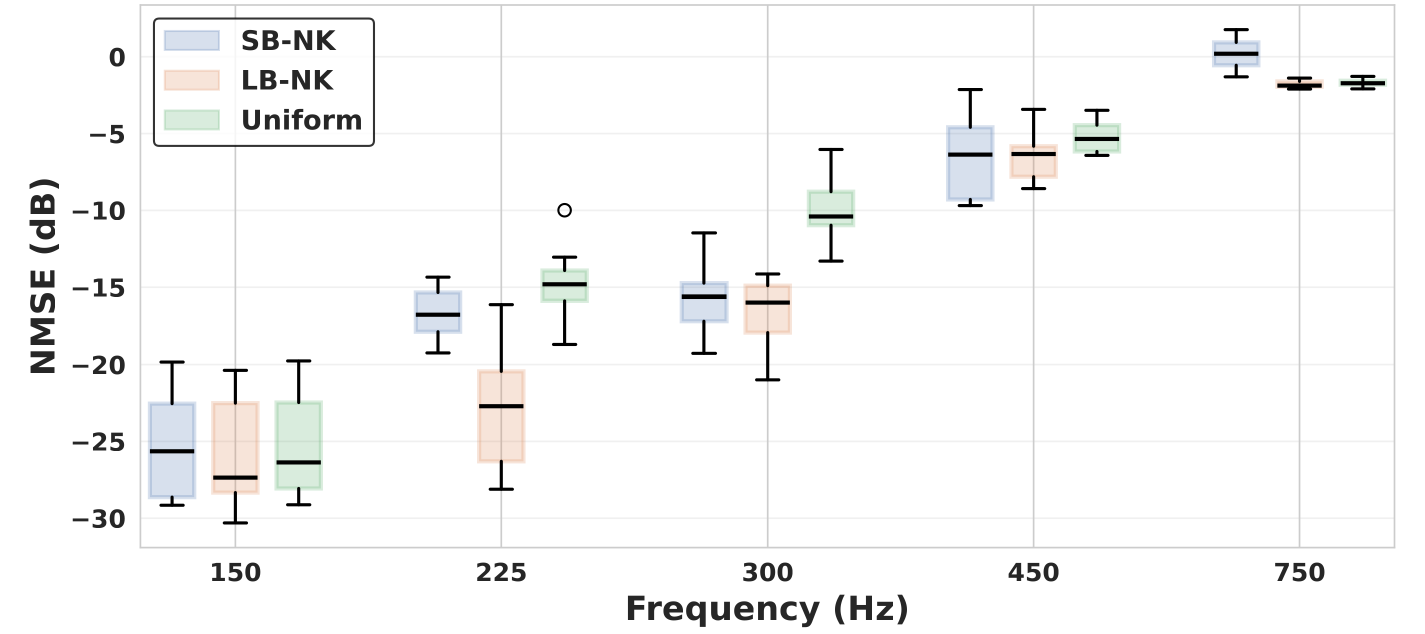}\\
    \caption{Medians, inter-quartile ranges, and outliers of NMSE are plotted as a boxplot.}
    \label{fig:nmse_18_boxplot}
\end{figure}

Fig.~\ref{fig:nmse_18_average} shows the NMSE averaged across all test sources as a function of frequency for a configuration with $M=18$ microphones. To provide a deeper insight into the statistical distribution of the error, Fig.~\ref{fig:nmse_18_boxplot} presents the corresponding boxplots at specific representative frequencies: $150$, $225$, $300$, $450$, and $750~\mathrm{Hz}$. As expected, all methods exhibit a monotonic degradation in performance as the frequency increases. This trend reflects the increasing spatial complexity of the acoustic field. Within the frequency range of $75$--$525~\mathrm{Hz}$, LB-NK consistently achieves a lower average NMSE than both SB-NK and the Uniform baseline. Furthermore, the tighter inter-quartile ranges observed at higher frequencies in the boxplots for LB-NK suggest that the multi-source training acts as an effective structural regularizer, leading to a more stable reconstruction that is less sensitive to the specific source realization, even under challenging high-frequency conditions.
 

\begin{figure}[t!]
    \centering
    \begin{subfigure}[t]{0.5\textwidth}
    \includegraphics[width=1.0\columnwidth]{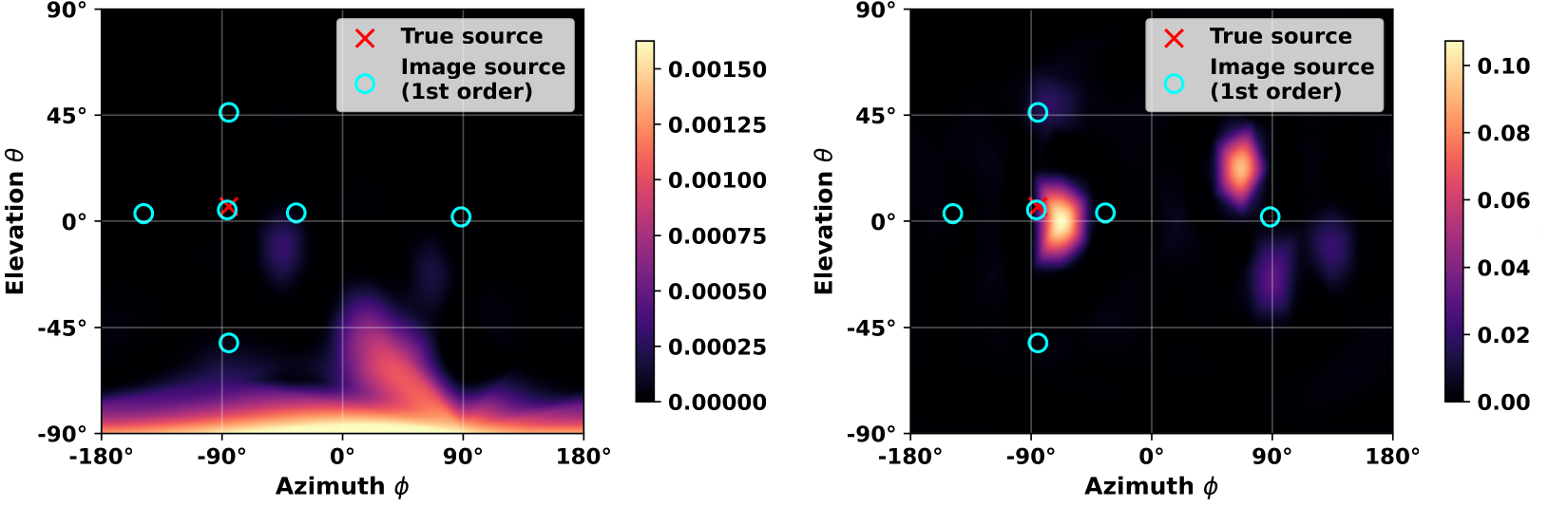}
    \caption{$150~\mathrm{Hz}$, $M=18$}
    \vspace{10pt}
    \end{subfigure}
    \begin{subfigure}[t]{0.5\textwidth}
    \includegraphics[width=1.0\columnwidth]{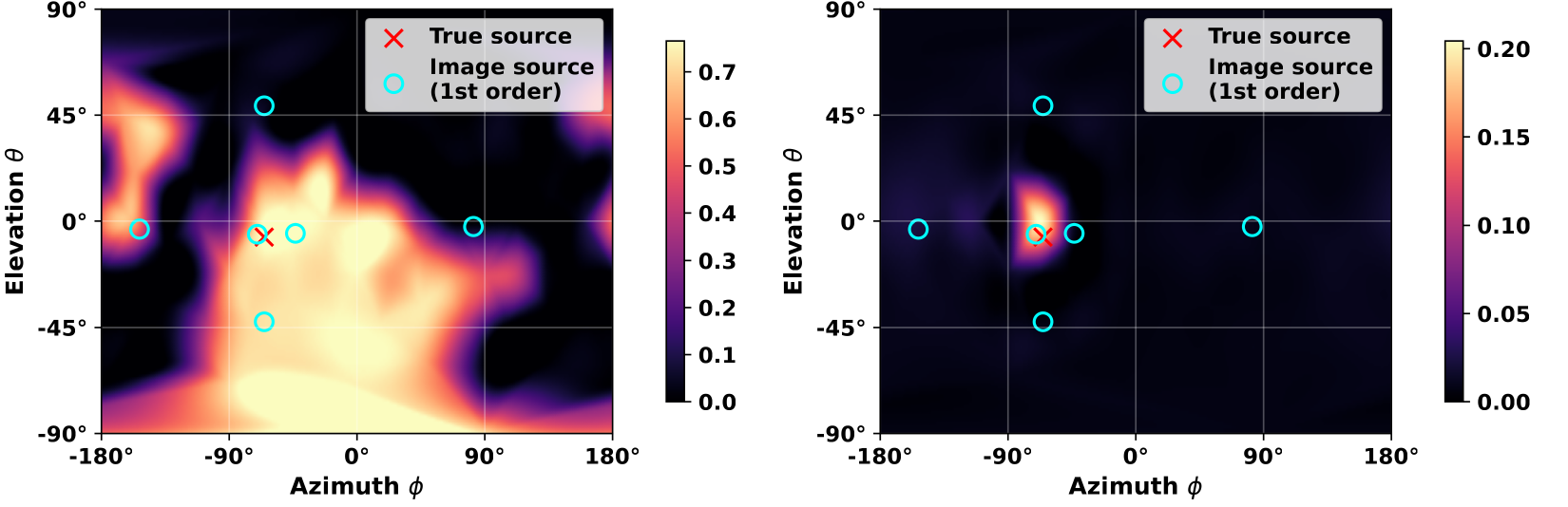}
    \caption{$1200~\mathrm{Hz}$, $M=50$}
    \end{subfigure}
    \caption{Directional weighting function obtained by SB-NK (left) and LB-NK (right). The directions of the true source and first-der image sources are indicated by red crosses and cyan circles, respectively.}
    \label{fig:heatmap}
\end{figure}

Fig.~\ref{fig:heatmap} shows the optimized directional weighting function $w$ for LB-NK and SB-NK at $150~\mathrm{Hz}$ and $1200~\mathrm{Hz}$. Whereas the directional weight of SB-NK tends to spread energy across multiple competing lobes with no clear alignment to the ground truth, the LB-NK weights are sharply concentrated on the directions corresponding to the direct source and the first-order early reflections. At lower frequencies, this concentration allows LB-NK to identify a physically consistent propagation direction, whereas SB-NK exhibits a more isotropic and less accurate distribution. As the frequency increases to $1200~\mathrm{Hz}$, the SB-NK fails to resolve a dominant directivity pattern, showing increased variance. In contrast, LB-NK maintains its focus near the true source and image-source locations. This behavior suggests that training across multiple sources acts as a powerful spatial regularizer, encouraging the network to learn a robust directional prior aligned with the underlying room geometry, rather than overfitting the specific angular noise or fluctuations of individual source realizations. 

\section{Conclusion}

We proposed a learning-based physics-constrained neural kernel for sound field estimation. The current method suffers from strong overfitting and poor generalization as the directional weighting function for the kernel function is optimized based solely on the single snapshot measurements. By constructing the directional weighting function using source-position-dependent INR, the kernel function based on the Herglotz wave function enables adaptation to the target acoustic environment using training ATF data while constraining the governing equation. We performed experimental evaluations to compare the proposed method with the snapshot-based methods, where the proposed method outperformed the snapshot-based method by properly capturing the directivity of the target sound field. Future work is to generalize this approach to multiple rooms. 

\bibliographystyle{IEEEbib}
\bibliography{str_def_abrv, refs, skoyamalab_en, refs_thesis}

\end{document}